# The Software Engineering Simulations Lab: Agentic AI for RE Quality Simulations

Henning Femmer[0000−0002−6059−4635], Ivan Esau

South Westphalia University of Applied Sciences (FH SWF),
Haldener Straße 182, 58095 Hagen, Germany,
lastname.firstname@fh-swf.de

**Abstract. Context and motivation.** Requirements Engineering (RE) quality still lacks empirical evidence on how specific requirement defects affect downstream activities. **Problem:** However, empirical data on the detailed effects of requirements quality defects is scarce, since it is costly to obtain. Furthermore, with the advent of AI-based development, the requirements quality factors may change: Requirements are no longer only consumed by humans, but increasingly also by AI agents, which might lead to a different efficient and effective requirements style. **Principal ideas:** We propose to extend the RE research toolbox with *Agentic AI simulations*, in which software engineering (SE) processes are replicated by standardized agents in qualitative simulations. We argue that their speed and simplicity makes them a valuable addition to RE research, although limitations in replicating human behavior need to be studied and understood. **Contribution:** This paper contributes a first concept, a research roadmap, a prototype, and a first feasibility study for RE simulations with agentic AI. Study results indicate that even a naïve implementation leads to executable simulations, encouraging technical improvements along with broader application in RE research.

## 1 Motivation

Most researchers in requirements engineering (RE) agree that "*requirements are a means to an end rather than an end in themselves*" [3, p.14], emphasizing that requirements quality is context-dependent and defined by how and by whom requirements are used. Formalizing this paradigm, the activity-based RE artifact quality model (ABRE-QM) allows us to precisely define quality in a falsifiable way as measurable and observable factors that have measurable consequences on requirements-affected activities [6, 8]. Theoretically, this allows to systematically evaluate in which contexts quality factors matter. However, looking at the state of the art [13], the vast number of factors and possible impacts exposes a fundamental problem: **Problem 1:** There is currently no solid and economically viable way to understand the impact of requirements quality and we have limited evidence to extrapolate from [13].

In addition, we see a new usage of requirements (in ABRE-QM terms): generative AI may change quality factors and impacts; we need to understand which requirements quality factors improve the output quality of AI-based tools (e.g. code-generators). Initial studies indicate that requirements quality matters [7], though perhaps differently than in human interaction [21]. Finally, the rapid evolution of AI models increases the problem further. **Problem 2:** Generative AI introduces new quality factors, since requirements now affect both human and automated consumers. And as AI models evolve, we need fast, cost-effective ways to study how requirement quality influences downstream processes.

**General Idea:** We argue for adding Agentic-AI-based Simulations to the toolbox of empirical RE researchers: For this, we define a software development lifecycle (SDLC) and replace the participating humans with a set of AI agents such as agentic developers or testers, who simulate typical project behavior. Since this is purely automatic, we can replicate the experiment multiple times and feed a subset of the replications with requirements that differ in quality. After several repetitions, we can compare the outcomes and thus understand the impact of the requirements quality factors on the AI agents.

We argue that, while surely coming with limitations that need to be studied in this stream of research, this method could be a fitting tool for the aforementioned problems due to its low cost and quick execution. Given that the limitations of such simulations are understood, we argue that researchers could build a more complete model of RE quality for both AI-centric and human-centric SE processes enabling better-informed decisions in the context of RE.

**Scope of this Work:** The focus of this paper is to describe the general idea and evaluate whether the approach is computationally feasible at the current state of agentic AI. Therefore, for this first feasibility study we operationalized requirements quality through a broad list of observable quality factors (i.e. complex sentence structure, incorrect legal binding, inconsistent terminology, passive voice, missing coherence, and technical density) and assessed whether we can see any impact on the produced output, here operationalized by merge success, unit test coverage and unit test passing, calculation time and resource usage. Given that feasibility can be shown, future studies can use the methodology to assess other quality factors and/or to assess the impact of these quality factors in detail.

## 2 Background and Related Work

Due to space constraints we cannot give an in-depth review into the various fields that converge in this work. Instead, this section summarizes the key conceptual foundations that informed our study.

**Requirements Quality** has been defined from multiple perspectives in the past. Standards such as ISO 29148 or the IREB CPRE Syllabus have postulated quality as a set of defined quality factors. The differences of quality factors in IREB and ISO 29148, e.g. as analyzed in [6], indicates that such a selection is subjective and requires a different theoretical approach if we want to systematically evaluate and empirically falsify them.

Formalizing this modern paradigm, the activity-based RE artifact quality model (ABRE-QM) [6, 8] defines quality factors as *facts* of the *requirements entity* that have *impacts* on effectiveness or efficiency of *requirements-affected activities* executed by (typically human) *agents*. For example in a use case document (entity), the presence of a complex sentence structure (entity fact) increases (impact) the time needed (activity fact) to understand the requirement (activity) for a tester (agent) under certain contexts [8]. In theory, this defines quality as falsifiable statements which can be validated and refined, leading to a holistic, empirically grounded quality model. But despite individual studies on the impact of selected quality factors, such as ambiguity [2, 15] or passive voice [5, 9, 11], systematic analysis [13] shows that, overall, we lack empirical evidence on the detailed impact of individual RE quality factors.

**Simulations** have long been used i.a. to analyze dynamic, complex systems [14]. Therefore, we rely on established guidelines and processes [14].

LLMs now play a major **role in RE and SE** [10, 4], producing useful artifacts for elicitation [18], quality assurance [12], modeling [7], code [23] and test generation [20], usability testing [16], and traceability [21]. However, to the best of our knowledge the existing research examines only activities in isolation and lacks an integrated view of the overall process.

Other works focus on **evaluating the impact of requirements quality on automated development.** Recent studies suggest that the linguistic and structural quality of requirements influences performance of automated development tools and agents [21]. High-quality requirements appear to lead to more consistent and complete outcomes in downstream tasks, whereas ambiguity and complexity can reduce reliability. Despite these first indications, systematic research remains scarce, and existing studies usually focus on one isolated activity.

## 3 Agentic Simulations in Requirements Engineering with DevOps Pipelines

### 3.1 Vision

As we envision the process, (1) the experimenter begins by setting up template (*baseline*) projects in the DevOps environment, for example, one baseline project describing requirements as tickets in active voice and one baseline project describing the same requirements as tickets in passive voice. (2) The experimenter opens the simulation system and defines, i.a., the number of project replications (*clones*) per baseline. (3) The simulation system then creates the requested number of identical clones of the baseline projects in the DevOps system and (4) starts a number of AI *agents* according to the defined parameters, which now take up different tasks, such as planning, developing, testing or reviewing. (5.1) These agents simulate the behaviors of actual project members and work directly on the DevOps platform. (5.2) Modern DevOps platforms contain automated quality checks through a CI/CD pipeline mechanism, which is run on any change of the codebase. Typical pipelines contain compiler checks, unit tests, and static code analysis but can be extended to any type of automated analysis. (5.3) The feedback from these pipelines now flows back to the agents, which helps them to iteratively solve their individual tasks (continues with step 5.1 until the feature is approved by the review agent). (6) After all features are implemented or certain exit criteria are fulfilled, experimenters can check the outcomes (e.g. the created code) and compare the clones, e.g., did the active voice requirements have a higher success rate, faster code generation with less agentic iterations, or less defects according to unit tests?

This would provide researchers with data on the question whether passive voice requirements are a relevant requirements quality factor for agentic SE projects. And in a second step, given that a deeper understanding of similarities and differences between human and agentic SE agents behavior allows us to extrapolate from agentic to human behavior, it might even provide first insights for improving human-driven SE projects.

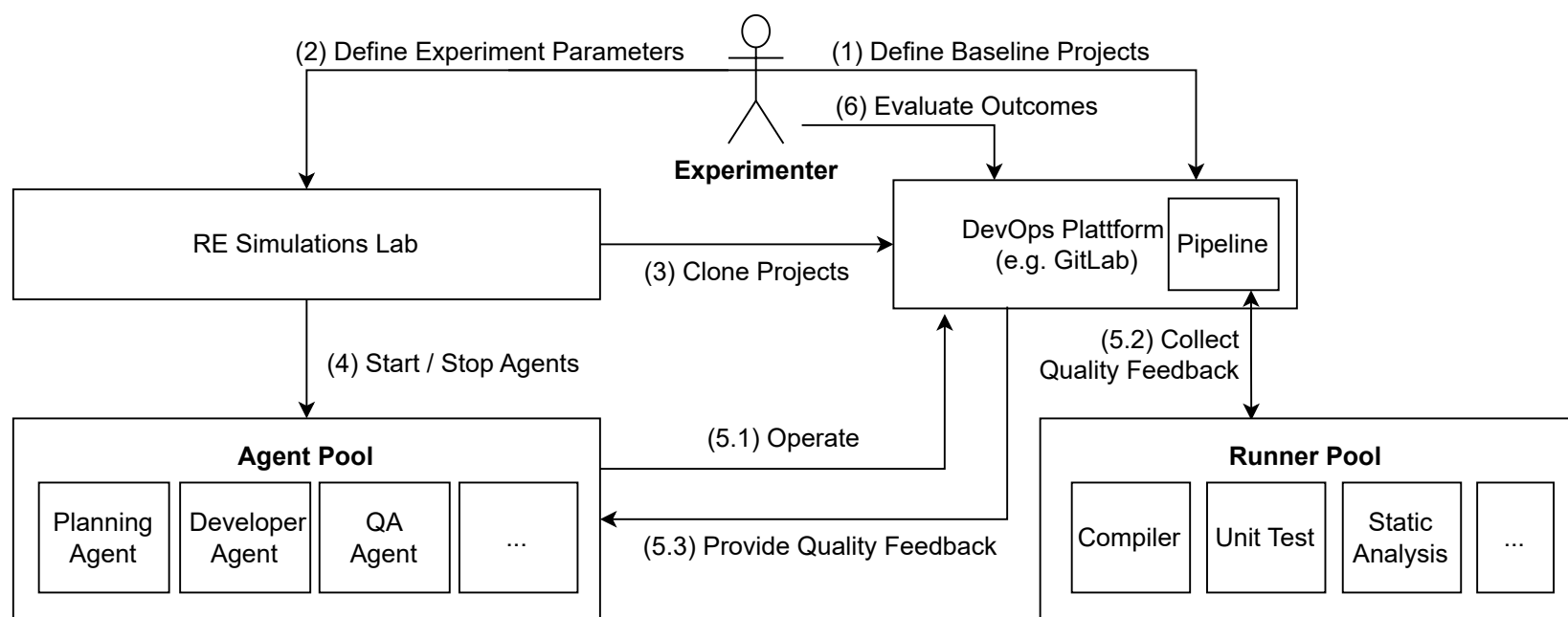

**Fig. 1.** The Process Behind A Pipeline-based Agentic Simulation

### 3.2 Research Roadmap

The usage of agentic simulations in RE has two research objectives (RO): First, to understand how to do RE effectively and efficiently for agentic AI. Second, to understand to what degree these results can be generalized to human-centric SE processes.Together, this provides practitioners with a more complete and evidence-based RE quality model and researchers with a novel research method to study certain RE phenomena in a simulated environment. The following roadmap describes our long-term research agenda.

### 3.3 RO 1: Define a RE Quality Model for AI-coding Agents

The first objective is to design and implement a technical infrastructure that replicates realistic SE processes on a DevOps platform such as GitLab. For this process, we need to define actors, artifacts and actions. For actors, we need to elicit a set of agent roles (such as developer or testers) and envision to use docker as a standardized interface to agents. For agent actions the Model Context Protocol (MCP) could form a standardized interface to the DevOps system.

**Research Method:** Since the quality model is a continuous endeavor, we should design the research accordingly. Therefore, we propose to apply a design-science approach with agentic simulations as validation methods. In each cycle, researchers will extend the quality model with further stakeholders, activities, quality factors, impacts and context.

**RQs:** RQ 1.1: Can agentic AI cooperatively implement requirements in a DevOps process simulation? RQ 1.2: What is the impact of known quality factors onto agents' activities? RQ 1.3: What is the cost (financial & environmental) of the simulations?

**Data Collection:** To answer these, we proceed as discussed in the previous section. The approach allows to collect various performance metrics for each clone from the DevOps environment, such as the number of successful merge requests, automated unit test metrics, manual assessment of requirements fulfillment, static QA assessments about the code output, and dynamic QA assessments such as usability tests. Both are elicited for the simulation results in general (RQ 1.1) and per quality factor (i.e. project A vs. B, RQ 1.2). Furthermore, we elicit token consumption, cost, and runtime metrics for the used LLMs (RQ 1.3).

**Data Analysis:** We calculate success percentage for RQ 1.1, comparative analysis for RQ 1.2 and average costs per replication for RQ 1.3.

### 3.4 RO 2: Understand agentic simulation's abilities and limitations in generalization to human behavior

The second RO investigates how well simulated processes represent human work.

**Research Method:** Comparative analysis of human and agentic AI behavior in controlled environments must be conducted in which human engineers replace one of the agents while the rest of the process remains automated.

**RQs:** RQ 2.1: How does the context taken into account by the humans differ from the agents? RQ 2.2: How does the process executed by the humans differ? RQ 2.3: How do the artifacts created by humans differ?

**Data Collection:** Quantitative and qualitative metrics as in RO1 as well as interview and observation data from the experiment.

**Data Analysis:** By comparing process metrics and output quality, differences in reasoning, error patterns, and decision strategies can be identified. This enables systematic assessment of where simulations produce representative results and where human-specific factors (such as interpretation, ambiguity resolution, or creativity) play a decisive role. This data will help determine under which conditions simulations are a valid proxy for human-driven software processes.

### 3.5 Risks

Any research involving LLMs risks evolving model behavior affecting reproducibility, data leaks, and varying outputs. We suggest mitigating this by using open-weights models (allowing version pinning) and publishing complete interaction logs as proposed in [19]. Besides these standard risks, this roadmap faces two specific risks:

**$R_1$: Limited abilities of individual agents.** The agents used in this study might not yet be able to handle complex or dependent tasks reliably, which would limit how realistic the simulations can be.

**$R_2$: Scalability problems due to high computing, financial or environmental costs.** Running many simulations could require substantial resources, which could call into question whether the results justify these costs.

## 4 Software Engineering Simulations Lab (SESL)

To understand the risks specified and get first insights into RO1 of this research agenda, we set up a feasibility study with a limited scope as defined in the introduction. Following the methodology of [14], the **purpose** of this work focuses on *process improvement* of RE processes by *understanding* the impact of quality factors in RE artifacts on subsequent manual and automated SE activities. The **scope** of the simulations is a single project or a single iteration thereof. The **reference behavior** is the real-world behavior of iterative SE projects in a DevOps/Git Flow context, working with or without agentic AI systems. In these iterations, automated CI/CD pipelines provide quality feedback. Our **model concepts** can be found in Fig. 1. The subsequent sections describe the remaining two steps of the simulation process according to [14].

### 4.1 Executable Model: SESL architecture

To evaluate the risks and to evaluate feasibility, we developed a prototype of SESL, including a naïve baseline implementation of all relevant components.

For our prototype, we used our GitLab instance with four GitLab pipeline runners. As a pipeline, we employed Java compilation and JUnit testing with JaCoCo for coverage. Feedback from static analysis was not included in our feasibility study to shorten iterations. The SDLC was operationalized through a simplification of an agile project in a typical Git Flow [17] style, defining agents by their I/O behavior (see Tab. 1). Agents were run in sequential orchestration in the form of the supervisor pattern [22]. In addition, each agent was instructed to create a textual report as a markdown file in a dedicated folder at the end of their action, which was read by the subsequent agents to reduce hallucinations by setting the context and enable human validation (as proposed by [1]). All further design details, such as applied prompts, source code etc. can be found in the supplementary material.

**Table 1.** Employed AI agents in feasibility study

| Agent | Responsibilities | ⇥**Inputs** & ↦**Outputs** |
|---|---|---|
| Planning Agent | Order work items and define plan | ⇥Work items<br>↦Prioritized plan, Architecture file |
| Coding Agent | Create code for a work item | ⇥Work items, Existing code, Architecture<br>↦Branches, Code commits |
| Testing Agent | Create test cases for a work item | ⇥Work items, Existing code, Architecture<br>⇥Branches, Commits with unit test code |
| Review Agent | Check pipeline and unit tests for success and code against requirements implementation | ⇥Pipeline results<br>↦Merge request |

### 4.2 Simulation Results: First Risk Insights from a Feasibility Study

As a first feasibility study, we ran a simulation using the aforementioned prototype architecture, looking into RQ 1.1-1.3.

**Data Collection and Analysis:** To evaluate and/or mitigate the aforementioned risks, we follow existing guidelines [19, 1] by recording model version and date, publishing all prompts and configurations (available in our repository), saving results in Git, and using DeepSeek as an open weights model. Future work will include comparing output consistency and validating results with human reviewers as recommended in [1].

As a baseline, we created two projects in GitLab with five requirements each for implementing a battleships game. One baseline (A) contains five work items which could be considered high quality requirements, the other (B) contains the same work items, but we injected six requirements quality defects: Complex sentence structure, incorrect legal binding, inconsistent terminology, passive voice, missing coherence, and technical density. We intentionally combined multiple defects to increase the effect for the feasibility study. Each work item was described using a title, a user story, a detailed description and three acceptance criteria each, with an average of 162.4 words in total for the whole description.

The SESL platform replicated both baselines 10 times each, leading to a total of 20 simulations. Of these, 6 simulations needed to be repeated a second time due to GitLab getting stuck. The agents were run using the DeepSeek-V3.2-Exp model on Oct. $10^{\text{th}}$-$22^{\text{nd}}$ 2025 at a temperature of 0.2 to reduce variance.

**First Results RQ 1.1&1.2:** The overall results are promising: With just a naïve baseline implementation, 34% of the requirements (A:38%, B:30%) were automatically implemented and merged, 23% (A:26%, B:20%) passed all unit tests provided by testing agents. The average line coverage in unit tests was 40% (A:44%, B:37%). A first inspection of the details indicates that the agents struggled particularly with the defects injected into requirement#2 (e.g. on average 18.5 points worse line coverage for agents working with defective requirements vs. defect-free requirements, 2.5 times more timeouts and errors during pipeline execution). Overall this provides first evidence for a higher success rate and higher quality for agents working with the high quality requirements (A).

**First Results RQ 1.3:** The simulation for these 20 successful projects took 110 hours or 5.5 hours per clone (A: 5.3h, B: 5.7h) with GitLab pipelines being the main bottleneck. Required tokens per clone averaged at 94.2 mio. input and 269.6k output tokens, leading to costs of $3.27 per simulation clone. According to a carbon calculation tool (`https://llmemissions.com`) the LLM component of the experiment had an environmental footprint of approx. 0.6 kg CO2 overall.

## 5 Discussion, Summary and Outlook

This work proposes understanding RE quality by simulating human behavior with agentic AI. These agents work directly on clones of DevOps projects and get feedback from CI/CD pipelines. The work furthermore contributes a prototypical architecture as well as first results from a feasibility study with 20 projects on a total of 100 requirements run sequentially over a period of 110 hours.

Our results indicate that running simulations with LLM-based agents is fast and cost-effective, and it can produce working software (RQ 1.1). Moreover, the early results suggest that lower requirements quality has a negative impact on the agents' output (RQ 1.2). We furthermore present first data on environmental and financial cost and execution time, supporting our claim that simulations could provide a cost- and time-effective method (RQ 1.3). Future work will analyze the results in detail, looking at various assumed quality factors, such as passive voice, complex language, structure, etc. To which degree the created results generalize to human behavior (RO2) is also out of scope of this preview.

*Threats to validity:* As this work presents an early research preview, threats to validity are many-fold. Results are constrained by the prototype setup, the selected AI model and instructions. Therefore, future work must include state-of-the-art agents and look deeper into addressing risks of overfitting. A qualitative analysis into the actions of the agents and a comparison against human participants is furthermore necessary. This includes also an analysis on the consistency and hallucinations of the agentic behavior within a model, over time and across models. The chosen requirements reflect a well-known game and are therefore probably part of the training set, which might lead to a lesser effect of the requirements quality, since the ambiguity can be compensated by the model. It is furthermore a small set of requirements, scalability beyond that remains unclear at this early point. The reported execution times are mostly constrained by the GitHub setup and our approach of sequential execution and are more of an upper bound for the given project. A follow-up detailed analysis will also report on the specific impact of other quality factors on other activity facts, such as manual assessment of the correctness of requirements implementation or test quality.

**Data Availability Statement** The code of our implementation can be found at `https://github.com/FH-SWF-SSQL/SESL`. The template projects as well as the resulting executed projects with all pipeline logs from the simulations are available on GitLab `https://gitlab.nibbler.fh-swf.de/publications/sesl-feasibility`.

## References

1. Baltes, S., et al.: Guidelines for empirical studies in software engineering involving large language models. arXiv preprint arXiv:2508.15503 (2025)
2. de Bruijn, F., Dekkers, H.L.: Ambiguity in natural language software requirements: A case study. In: REFSQ (2010)
3. Bühne, S., et al.: CPRE Foundation Level - Syllabus v.3.2.0 (2024)
4. Dabrowski, J., Cai, W., Bennaceur, A., Nuseibeh, B., Alrimawi, F.: Intelligent agents for requirements engineering: Use, feasibility and evaluation. In: RE (2025)
5. Femmer, H., Kučera, J., Vetrò, A.: On the impact of passive voice requirements on domain modelling. In: ESEM (2014)
6. Femmer, H., Vogelsang, A.: Requirements quality is quality in use. IEEE Software **36**(3) (2018)
7. Ferrari, A., Abualhaija, S., Arora, C.: Model generation with llms: From requirements to uml sequence diagrams. In: RE Workshops (REW). IEEE (2024)
8. Frattini, J., Montgomery, L., Fischbach, J., Mendez, D., Fucci, D., Unterkalmsteiner, M.: Requirements quality research: a harmonized theory, evaluation, and roadmap. Requirements engineering **28**(4) (2023)
9. Frattini, J., et al.: A second look at the impact of passive voice requirements on domain modeling: Bayesian reanalysis of an experiment. In: WSESE (2024)
10. Hou, X., et al.: Large language models for software engineering: A systematic literature review. Transactions on Software Engineering and Methodology **33**(8) (2024)
11. Krisch, J., Houdek, F.: The myth of bad passive voice and weak words an empirical investigation in the automotive industry. In: RE (2015)
12. Lubos, S., Felfernig, A., Tran, T., Garber, D., El Mansi, M., Erdeniz, S.P., Le, V.: Leveraging llms for the quality assurance of software requirements. In: RE (2024)
13. Montgomery, L., et al.: Empirical research on requirements quality: a systematic mapping study. Requirements Engineering **27**(2) (2022)
14. Müller, M., Pfahl, D.: Simulation methods. In: Guide to advanced empirical software engineering. Springer (2008)
15. Philippo, E.J., et al.: Requirement ambiguity not as important as expected - results of an empirical evaluation. In: REFSQ (2013)
16. Pourasad, A.E., Maalej, W.: Does GenAI Make Usability Testing Obsolete? In: ICSE (2025)
17. Ríos, J.C.C., Embury, S.M., Eraslan, S.: A unifying framework for the systematic analysis of git workflows. Information and Software Technology **145** (2022)
18. Ronanki, K., Berger, C., Horkoff, J.: Investigating chatgpts potential to assist in requirements elicitation processes. In: SEAA (2023)
19. Sallou, J., Durieux, T., Panichella, A.: Breaking the silence: the threats of using llms in software engineering. In: ICSE NIER (2024)
20. Steenhoek, B., et al.: Reinforcement learning from automatic feedback for high-quality unit test generation. In: DeepTest (2025)
21. Vogelsang, A., Korn, A., Broccia, G., Ferrari, A., Fischbach, J., Arora, C.: On the impact of requirements smells in prompts: The case of automated traceability. In: ICSE-NIER (2025)
22. Wang, Y., et al.: Agents in software engineering: Survey, landscape, and vision. Automated Software Engineering **32**(2) (2025)
23. Wei, B.: Requirements are all you need: From requirements to code with llms. In: RE (2024)